\documentclass[a4paper,11pt]{article}
\pdfoutput=1
\usepackage{jcappub}

\usepackage{color}
\usepackage{amsfonts}
\usepackage{graphicx}
\usepackage{amssymb}

\usepackage{eufrak}
\usepackage{mathrsfs}

\usepackage{mathtools}
\usepackage{ulem}
\usepackage{bm}
\usepackage{xcolor}

\renewcommand{\thesection}{\arabic{section}}
\renewcommand{\theequation}{\thesection.\arabic{equation}}

\def\laq{~\raise 0.4ex\hbox{$<$}\kern -0.8em\lower 0.62ex\hbox{$\sim$}~}
\def\gaq{~\raise 0.4ex\hbox{$>$}\kern -0.7em\lower 0.62ex\hbox{$\sim$}~}

\def\beq{\begin{equation}}
\def\eeq{\end{equation}}
\def\bea{\begin{eqnarray}}
\def\eea{\end{eqnarray}}

\def \pa {\partial}
\def \ra {\rightarrow}
\def \ti {\widetilde}
\def \la {\lambda}

\def \lp {\lambda_{\rm P}}

\def \La {\Lambda}

\def \da {\delta}
\def \b {\beta}
\def \a {\alpha}

\def \Ga {\Gamma}
\def \ga {\gamma}
\def \sg {\sigma}

\def \da {\delta}

\def \r {\rho}

\def \Om {\Omega}

\def \L {{\cal L}_m}
\def \og {\overline g}

\title{Linearized propagation equations \\ for metric fluctuations in a general (non-vacuum) background geometry}

\author[a]{G. Fanizza,}
\author[b]{M. Gasperini,}
\author[b]{E. Pavone}
\author[b]{and L. Tedesco}

\affiliation[a]{Instituto de Astrofis\'ica e Ci\^encias do Espa\c{c}o,
Faculdade de Ci\^encias da Universidade de Lisboa,
Edificio C8, Campo Grande, P-1740-016, Lisbon, Portugal}
\affiliation[b]{Dipartimento di Fisica, Universit\`a di Bari, 
Via G. Amendola 173, 70126 Bari, Italy,\\
and Istituto Nazionale di Fisica Nucleare, Sezione di Bari, Italy}

\emailAdd{gfanizza@fc.ul.pt}
\emailAdd{maurizio.gasperini@ba.infn.it}
\emailAdd{~~~~~~~~~~e.pavone2@studenti.uniba.it}
\emailAdd{luigi.tedesco@ba.infn.it}

\abstract{The linearized dynamical equation for metric perturbations in a fully general, non-vacuum, background geometry is obtained from the Hamilton variational principle applied to the action up to second order. We specialize our results to the case of traceless and transverse metric fluctuations, and we discuss how the intrinsic properties of the matter stress tensor can affect (and modify) the process of gravity wave propagation even in most conventional geometric scenarios, like (for instance) those described by a FLRW metric background. We provide explicit examples for fluid, scalar field and electromagnetic field sources.}

\keywords{Gravitational waves; metric perturbations; gravitational theory; cosmology.
 
\vskip13pt plus8pt minus11pt

\noindent{\bfseries\large\sffamily{Preprint:}} BA-TH/804-20 
}

\begin{document}

\maketitle


\section{Introduction}
\label{Sec1}
\setcounter{equation}{0}

The dynamics of tensor metric perturbations in non-trivial geometric backgrounds is a crucial ingredient in the process of amplification of the vacuum fluctuations,  and in the related production of a relic background of cosmic gravitational waves (GW) (see e.g.  \cite{Muk}). The aim of this paper is to derive such dynamics from an action variational principle for the most general type of  background geometry, sourced by arbitrarily given matter-energy distributions. 

What we need, to this purpose, is to apply to the action the Hamilton variational formalism expanded up to second order. We will assume that the gravitational interaction is described by the usual Einstein-Hilbert action, but the same procedure can be easily applied to more general (e.g. higher-derivative, non-minimally coupled, scalar-tensor, etc) models of gravitational dynamics.

The first-order variation of the Einstein action with respect to the metric generates, as is well known, non-vanishing boundary terms that are to be cancelled by adding the variational contributions of an appropriate (York-Gibbons-Hawking) boundary action, in order to recover the background Einstein equations. At the second variational order, on the contrary, it will be shown that all  boundary terms are automatically vanishing, that the condition of stationary action leads to a dynamical equation for the evolution of the metric fluctuations, and that such equation exactly corresponds to the linear perturbation of the background gravitational equations.

The main result of this paper is that we may expect in principle interesting changes to the dynamics of GW propagation in vacuum (see e.g. \cite{Mis}), even without considering modified theories of gravity. Even for the Einstein theory of gravity, and for gravitational sources minimally coupled to the geometry of a Riemann space-time manifold, we find that non-trivial contributions to the GW propagation equation are possible, in principle, provided the the energy-momentum of the matter sources  satisfies some special appropriate condition. It should be stressed, also, that we are perturbing {\it only} the metric tensor appearing in the coupling of the matter fields to gravity, and not the other physical variables (energy-density, velocity, \dots) typical of the matter fields themselves.

The paper is organized as follows. In Sect. \ref{sec2} we derive the most general evolution equation for linear metric perturbations by imposing on the full action to be stationary, up to second order, with respect to the variation of the metric tensor. In Sect. \ref{sec3} we concentrate our discussion on the case of transverse and traceless metric fluctuations and on the general class of non-vacuum background spacetimes compatible with this choice. We give examples with scalar fields and fluids as gravitational sources, and we discuss how the fluid viscosity can affect GW propagation even in a very simple, spatially flat, FLRW background geometry. In Sect. \ref{sec4} we presents our final remarks, and suggest possible applications of the results derived in this paper. Finally, the Appendix A contains 
 a few explicit computations  whose results are used in the main text; in the Appendix B we report a further example of modified GW propagation in a background geometry with the electromagnetic field as the dominant gravitational source.

{Conventions}: in this paper we will use the metric signature $g_{\mu\nu}= {\rm diag} (+,-,-,-)$;  the Riemann tensor is defined as $R_{\mu\nu\a}\,^\b = \pa_\mu \Ga_{\nu\a}\,^\b + \Ga_{\mu\r}\,^\b \Ga_{\nu\a}\,^\r - \cdots$, the Ricci tensor is given by $R_{\nu\a}= R_{\mu\nu\a}\,^\mu$, and we define covariant derivatives with the following convention: $\nabla_\mu A_{\nu \cdots}= \pa_\mu A_{\nu  \cdots} - \Ga_{\mu\nu}\,^\a A_{\a  \cdots}+  \cdots$. 


\section{Evolution equation for linear metric perturbations in a general background geometry}
\label{sec2}
\setcounter{equation}{0}

Let us start by linearly expanding the metric tensor around a given background metric $g_{\mu\nu}$, by defining  
\beq
g_{\mu\nu} \ra \og_{\mu\nu} = g_{\mu\nu} + \da g_{\mu\nu},
\label{21}
\eeq
and let us impose on the action $S$ (describing the geometry and its interactions with the matter fields), to be stationary ($\da S=0$) with respect to the above background perturbation. By applying the standard variational formalism to first order in $\da g_{\mu\nu}\equiv h_{\mu\nu}$ we obtain, as is well known, the  field equations for the background metric $g_{\mu\nu}$. To second order,   the condition of stationary action gives a differential equation governing the dynamical evolution of the metric fluctuations  $h_{\mu\nu}$, and exactly corresponding -- as will be shown -- to the linear perturbation of the background gravitational equations. 

In order to compute the action variation induced by the infinitesimal transformation (\ref{21}), to any given order in the powers of $\da g_{\mu\nu}$, it is convenient to apply the formalism of functional differentiation, and expand the action in Taylor series of functional derivatives with respect to the metric. Let us recall, to this purpose, that if we have an arbitrary function of the metric and of its derivatives, $A= A(g, \pa g, \cdots)$, and we perform the metric expansion (\ref{21}), we can expand $A(\og, \pa \og)$ in power series of $\da g_{\mu\nu}$ as follows:
\beq
A(\og, \pa \og) = A(g, \pa g) + \da^{(1)}A + \da^{(2)}A + \cdots
\label{22}
\eeq
where
\bea
\da^{(1)}A &=& \left(\da A\over \da \og_{\mu\nu}\right)_0 \da g_{\mu\nu}, 
\label{23} \\
\da^{(2)}A &=& {1\over 2} \left({\da \over \da \og_{\r\sg}}{\da A \over \da \og_{\mu\nu}}\right)_0 \da g_{\r\sg} \,\da g_{\mu\nu},
\label{24}
\eea
and where the subscript $``0"$ denotes a function to be evaluated for the unperturbed metric, i.e. for $\og_{\mu\nu}=g_{\mu\nu}$.
Finally, we have introduced  the symbol of functional derivative, $\da / \da g_{\mu\nu}$, defined as usual by
\beq
{\da \over \da g_{\mu\nu}} A\left(g, \pa g, \cdots \right) \da g_{\mu\nu} \equiv
{\pa A\over \pa g_{\mu\nu}} \da g_{\mu\nu} +{\pa A\over \pa(\pa_\la g_{\mu\nu})} \pa_\la \da g_{\mu\nu}+ \cdots
\label{25}
\eeq
(we have used the commutation property $\da \pa_\la g_{\mu\nu}\equiv \pa_\la \og_{\mu\nu}- \pa_\la g_{\mu\nu} \equiv \pa_\la \da g_{\mu\nu}$, due to the fact that the metric transformation (\ref{21}) is performed {\it locally}, i.e. at fixed spacetime position).

As a simple illustrative example of the above formalism we may consider the expansion of the controvariant components of the metric tensor. By setting $A=g^{\a\b}$, using the definitions (\ref{23}) and (\ref{24}), and the property $d g^{\a\b} g_{\b\nu}= - g^{\a\b}d g_{\b\nu}$, we have: 
\bea
\da^{(1)} g^{\a\b} &=&\left(\pa\og^{\a\b} \over \pa \og_{\mu\nu}\right)_0 \da g_{\mu\nu} = - g^{\a\mu} g^{\b\nu} \da g_{\mu\nu}=-h^{\a\b},
\label{26} \\
\da^{(2)} g^{\a\b} &=& {1\over 2} \left({\da \over \da \og_{\r\sg}}{\da \og^{\a\b}\over \da \og_{\mu\nu}}\right)_0 \da g_{\r\sg} \,\da g_{\mu\nu}
\nonumber \\
&=&{1\over 2}
\left(g^{\a\r}g^{\mu\sg} g^{\b\nu}+g^{\a\mu}g^{\b\r} g^{\nu\sg}\right)\da g_{\r\sg} \,\da g_{\mu\nu}= h^{\a\mu}h_\mu\,^\b.
\label{27}
\eea
The functional expansion of $\og^{\a\b}$ to order $h^2$ then takes  the explicit form
\beq
\og^{\a\b} = g^{\a\b}- h^{\a\b} +h^{\a\mu}h_\mu\,^\b + \cdots,
\label{28}
\eeq
in obvious agreement with the condition
\beq
\og^{\a\b}\,\og_{\b\nu} = \da^\a_\nu + {\cal O}(h^3).
\label{29}
\eeq

Let us now apply the above formalism to compute the variation of the action for gravity and its matter sources, under the effects of the  transformation (\ref{21}), to the linear and to the quadratic order in powers of the metric fluctuation $\da g_{\mu\nu}$. Let us consider the Einstein model of gravity,  described by the action
\beq
S= \int d^4 x \left( -{1\over 2\lp^2} \sqrt{-g}\, R + \sqrt{-g}\,\L\right),
\label{210}
\eeq
where $\lp^2=8\pi G$ is the  Planck length parameter, $\L$ the Lagrangian density of the matter sources, and let us expand the infinitesimal variation of the action as follows:
\bea
&&
~~~~~~~~~~~~~~~~~~~~~~~~~~~~~~~~~~~
\da S \equiv \da^{(1)} S+\da^{(2)} S + \cdots =
\nonumber \\ &&
\!\!\!\!\! \!\!\!\!\!  \!\!\!\!\!   =
{1\over 2 \lp^2} \int d^4 x\left[ \da^{(1)}\left(- \sqrt{-g} R + 2 \lp^2 \sqrt{-g}\L\right)+  \da^{(2)}\left(- \sqrt{-g} R + 2 \lp^2 \sqrt{-g}\L\right)+ \cdots \right]. 
\label{211}
\eea

\subsection{First-order variation of the action}
\label{sec21}

The application to the above action of the variational principle to first order in $\da g_{\mu\nu}$ is well known, but it may be useful to briefly report here the results for its subsequent second-order generalization. 

Let us start with the standard definition of the dynamical energy-momentum tensor $T_{\mu\nu}$, given by
\beq
{\da \sqrt{-g} \,\L \over \da g_{\mu\nu}}= -{1\over 2} \sqrt{-g} \,T^{\mu\nu}.
\label{212}
\eeq
Let us write the Einstein action in terms of the Ricci tensor as $ \sqrt{-g}\,R= \sqrt{-g}\, g^{\a\b} R_{\a\b}$, and recall the well known result
\beq
d \sqrt{-g}= {1\over 2} \sqrt{-g} \, g^{\mu\nu} d g_{\mu\nu}= -  {1\over 2} \sqrt{-g} \, g_{\mu\nu}d g^{\mu\nu}.
\label{213} 
\eeq
We then easily find that the condition of stationary action to first order in $\da g_{\mu\nu}$, namely $\da^{(1)} S=0$, implies:
\beq
\da^{(1)} S={1\over 2 \lp^2} \int d^4 x \sqrt{-g}\left[ \left( R^{\mu\nu}-{1\over 2} g^{\mu\nu} R- \lp^2 T^{\mu\nu} \right) \da g_{\mu\nu} -
g^{\a\b} \da^{(1)} R_{\a\b} \right] =0,
\label{214}
\eeq
where 
\beq
 \da^{(1)} R_{\a\b} =\left( \da R_{\a\b} \over \da \og_{\mu\nu} \right)_0 \da g_{\mu\nu}.
\label{215}
\eeq
Let us  check now that this last contribution to the action integral represents the well known (non-vanishing) boundary term, to be cancelled by the variational contribution of an appropriate boundary action \cite{York,Gib} which must be added to the Einstein action. 

This important aspect of the variational formalism is of course largely understood and well illustrated in the specialistic literature on gravitational theory (and not only in the context of General Relativity, but also for generalized theories of gravity, see e.g. \cite{Dyer, Cosmai}).  For later use in this paper, it will be enough to recall here that the first functional derivative (with respect to the metric) of the Ricci tensor  can be expressed in terms of the so-called ``contracted Palatini identity" (see e.g. \cite{Gasp}), namely as
\beq
 \da^{(1)} R_{\nu\a}= \nabla_\mu \left( \da^{(1)} \Ga_{\nu\a}\,^\mu \right) 
 -  \nabla_\nu \left( \da^{(1)} \Ga_{\mu\a}\,^\mu \right).
 \label{216}
 \eeq
Here the covariant derivative $\nabla_\mu$ is performed in terms the background  metric $g_{\mu\nu}$, and $\da^{(1)} \Ga_{\nu\a}\,^\mu$ is the first-order perturbation (computed as a first functional derivative) of the Christoffel connection. By applying the metric property of the Riemann geometry ($\nabla_\mu g^{\a\nu}=0$), and using the Gauss theorem, we then find that the last contribution to Eq. (\ref{214}) can be written as an integral over the boundary hypersurface $\pa \Om$ of the considered space-time domain $\Om$: 
\bea
\!\!\!\!
\int_{\Om} d^4x \sqrt{-g}\,g^{\a\b} \da^{(1)} R_{\a\b}&=&
\int_{\Om} d^4x \sqrt{-g}\, \nabla_\mu \left( g^{\nu\a} \da^{(1)} \Ga_{\nu\a}\,^\mu- g^{\mu\a} \da^{(1)} \Ga_{\nu\a}\,^\nu \right)=
\nonumber \\ 
 &=&
\int_{\pa \Om} dS_\mu \sqrt{-g}\left( g^{\nu\a} \da^{(1)} \Ga_{\nu\a}\,^\mu- g^{\mu\a} \da^{(1)} \Ga_{\nu\a}\,^\nu \right).
\label{217}
\eea

On the other hand, an explicit computation of  $\da^{(1)} \Ga_{\nu\a}\,^\mu$, performed according to the definitions (\ref{23}) and (\ref{25}), provides a result which can be written in compact form as follows:
\bea
\da^{(1)} \Ga_{\nu\a}\,^\mu&=&-{1\over 2} \da g^{\mu\b} \left( \pa_\nu g_{\a\b}
+\pa_\a g_{\nu\b}-\pa_\b g_{\nu\a} \right)
+{1\over 2}  g^{\mu\b} \left( \pa_\nu \da g_{\a\b}
+\pa_\a \da g_{\nu\b}-\pa_\b \da g_{\nu\a} \right)
\nonumber \\ 
&=& {1\over 2}  g^{\mu\b} \left( \nabla_\nu h_{\a\b}
+\nabla_\a h_{\nu\b}-\nabla_\b h_{\nu\a} \right).
\label{218}
\eea
By inserting the above result into the boundary integral (\ref{217}) we can immediately check that, besides the terms proportional to the metric perturbation $\da g$ (whose contribution on the boundary is identically vanishing by assumptions, thanks to the rules of the Hamilton variational principle), there are also terms proportional to the derivatives of the metric variation, $\pa \da g$. Those terms  are {not} automatically vanishing on the boundary, and -- as already stressed -- they are to be eliminated by the addition of an appropriate boundary action. By including such an action into the variational procedure
%
 the overall contribution of the last term disappears from the integral (\ref{214}), and the condition of stationary action, to the first order in $\da g_{\mu\nu}$, leads to the well-known background equations for the unperturbed metric $g_{\mu\nu}$:
\beq
 R^{\mu\nu}-{1\over 2} g^{\mu\nu} R= \lp^2 T^{\mu\nu}.
 \label{219}
 \eeq
 
\subsection{Second-order variation of the action}
\label{sec22}
 
Let us now impose on the action to be stationary with respect to the infinitesimal transformation (\ref{21}) up to terms quadratic in $\da g_{\mu\nu}$. From Eq. (\ref{211}), and from the definition (\ref{24}), we are led to the condition
\beq
\da^{(2)} S\equiv {1\over 4 \lp^2} \int d^4 x \left[ {\da \over \da \og_{\r\sg}}{\da \over \da \og_{\mu\nu}}\left(- \sqrt{-g} \,R + 2 \lp^2 \sqrt{-g}\,\L\right) \right]_0 \da g_{\r\sg} \da g_{\mu\nu} =0.
\label{220}
\eeq
It is important to note  that there is no need  of explicitly including  into the above equation the additional (York-Gibbons-Hawking) boundary contribution, because the  second functional derivative of such term is identically vanishing. 
By exploiting the results already reported in the previous subsection, and using in particular Eq. (\ref{214}), we can then explicitly write the second order variation of the action as follows:
\bea
\da^{(2)} S&=& {1\over 4 \lp^2} \int d^4 x \left[ \da^{(1)} \left(\sqrt{-g}\, R^{\mu\nu}-
{1\over 2} \sqrt{-g}\, g^{\mu\nu} R\right) - \lp^2  \da^{(1)} \left(\sqrt{-g}\, T^{\mu\nu}\right) \right] \da g_{\mu\nu}
\nonumber \\
&-&  {1\over 4 \lp^2} \int d^4 x \left[ \da^{(1)} \left(\sqrt{-g}\,g^{\a\b}\right) \da^{(1)} R_{\a\b} + 2 \sqrt{-g}\,g^{\a\b} \da^{(2)} R_{\a\b} \right]
\label{221}
\eea
(the variational operators  $\da^{(1)}$ and $\da^{(2)}$ are defined, respectively, by Eqs. (\ref{23}), (\ref{24})). 

Let us now separately consider the two  integral terms $I_1$ and $I_2$ appearing in the second line of the above equation, namely
\beq
I_1=  \int d^4 x \da^{(1)} \left(\sqrt{-g}\,g^{\a\b}\right) \da^{(1)} R_{\a\b}, 
~~~~~~~~~~~
I_2= 2 \int d^4x \sqrt{-g}\, g^{\nu\a} \da^{(2)} R_{\nu\a},
\label{222}
\eeq
and show that, in spite of their non-trivial value, they give no contribution to the second-order variation of the action because they exactly cancel each other. 

Let us start with $I_1$. By recalling the results  (\ref{216}) e (\ref{218}) we can explicitly write $\da^{(1)} R_{\nu\a}$ as
\beq
\da^{(1)} R_{\nu\a}={1\over 2} \left( \nabla_\mu \nabla_\nu h_\a\,^\mu + \nabla_\mu \nabla_\a h_\nu\,^\mu - \nabla^2 h_{\nu\a} - \nabla_\nu \nabla_\a h \right),
\label{223}
\eeq
where we have set $h= g^{\mu\nu} \da g_{\mu\nu} = g^{\mu\nu} h_{\mu\nu}$, and we have denoted with $\nabla^2$ the covariant D'Alembert operator, $\nabla^2\equiv g^{\mu\b} \nabla_\mu \nabla_\b$. By using Eqs. (\ref{26}), (\ref{213}), (\ref{223}), by integrating by part, and neglecting the total divergences, we obtain:
\bea
\!\!\!\!\!\!\!\!\!\!
I_1&=&
\int d^4 x \sqrt{-g} \left( {1\over 2} h g^{\nu\a} - h^{\nu\a} \right)  \da^{(1)} R_{\nu\a}
\nonumber \\ &=& 
\int d^4 x \sqrt{-g}\left(-{1\over 2} \nabla_\mu h^{\nu\a} \nabla^\mu h_{\nu\a}
+\nabla_\mu h^{\nu\a} \nabla^\nu h_{\a}\,^\mu +{1\over 2} \nabla_\nu h \nabla^\nu h -  \nabla_\a h^{\a\mu}\nabla_\mu h \right). 
\label{224}
\eea
We have neglected the contribution of the total divergences because, using the Gauss theorem, they lead to integrals defined over the boundary hypersurface $\pa \Om$ with an argument proportional to $h_{\mu\nu}= \da g_{\mu\nu}$: hence, they are automatically vanishing thanks to the assumptions of the standard variational principle.

Let us now compute $I_2$. To this purpose, we need the second functional derivative $\da^{(2)} R_{\nu\a}$ which, by using the general definition of the Ricci tensor and of the covariant derivative with respect to the background metric, can be written as:
\beq
 \da^{(2)} R_{\nu\a}= \nabla_\mu \left( \da^{(2)} \Ga_{\nu\a}\,^\mu \right) 
 -  \nabla_\nu \left( \da^{(2)} \Ga_{\mu\a}\,^\mu \right) 
+ \da^{(1)} \Ga_{\mu\r}\,^\mu \da^{(1)} \Ga_{\nu\a}\,^\r
-\da^{(1)} \Ga_{\nu\r}\,^\mu \da^{(1)} \Ga_{\mu\a}\,^\r
 \label{225}
 \eeq
(see Appendix A). The first-order perturbation of the connection is given in Eq. (\ref{218}). At the second order we have
\beq
\da^{(2)} \Ga_{\nu\a}\,^\mu =
{1\over 2} \left({\da \over \da \og_{\ga\da}}{\da  \over \da \og_{\r\sg}}
\Ga_{\nu\a}\,^\mu 
\right)_0 \da g_{\ga\da}\, \da g_{\r\sg}, 
\label{226}
\eeq
and an explicit computation (see Appendix A) gives
\bea
\da^{(2)} \Ga_{\nu\a}\,^\mu&=&{1\over 2} h^{\mu\la}h_\la\,^\b \left( \pa_\nu g_{\a\b}+\pa_\a g_{\nu\b}-\pa_\b g_{\nu\a} \right)
-{1\over 2}  h^{\mu\b} \left( \pa_\nu h_{\a\b}
+\pa_\a h_{\nu\b}-\pa_\b h_{\nu\a} \right)
\nonumber \\
&\equiv& -{1\over 2}  h^{\mu\b} \left( \nabla_\nu h_{\a\b}
+\nabla_\a h_{\nu\b}-\nabla_\b h_{\nu\a} \right).
\label{227}
\eea

By inserting these results into the integral $I_2$ we then find that the  covariant-derivative terms appearing in the explicit expression of $\da^{(2)} R_{\nu\a}$ lead to boundary integrals proportional to $\da^{(2)} \Ga$.  In that case, the integrand is always proportional to the metric perturbation $\da g$, so that the corresponding  contribution to the action variation is automatically vanishing, as imposed by the  assumptions of the Hamilton variational principle. 
We are thus left with the contribution of the last two terms of Eq. (\ref{225}). By using the result  (\ref{218}) for $\da^{(1)} \Ga$, and combining all terms, we obtain:
\beq
I_2=
\int d^4 x \sqrt{-g}\, \left( {1\over 2} \nabla_\mu h^{\a\b} \nabla^\mu h_{\a\b} - {1\over 2} \nabla_\nu h \nabla^\nu h + \nabla_\mu h \nabla_\nu h^{\nu\mu} - \nabla_\nu h_\a\,^\mu \nabla_\mu h^{\nu\a}
\right).
\label{228}
\eeq

By comparing this result with Eq. (\ref{224}) we can immediately check that $I_1+I_2=0$, so that their contribution of the second-order action variation (\ref{221}) completely disappears. Hence, the condition of stationary action $\da^{(2)} S=0$ simply implies the vanishing of the first integral of Eq. (\ref{221}). Such a condition, by putting $R^{\mu\nu}= g^{\mu\a} g^{\nu\b}R_{\a\b}$ and $T^{\mu\nu} =   g^{\mu\a} g^{\nu\b}T_{\a\b}$, by factorizing the variational contribution $\da^{(1)} (\sqrt{-g} g^{\mu\a} g^{\nu\b})$, and imposing the validity of the background equations (\ref{219}), can be finally expressed as
\beq
 \da^{(1)} \left(R_{\a\b}-
{1\over 2}  g_{\a\b} R\right)= \lp^2 \, \da^{(1)} T_{\a\b}.
\label{229}
\eeq
It exactly corresponds to the condition one would obtain by computing the first functional differentiation of the background equations (\ref{219}). 

More explicitly, the above equation can be rewritten as
\beq
 \da^{(1)} R_{\a\b}-{1\over 2} R h_{\a\b}+{1\over 2} g_{\a\b} h^{\mu\nu} R_{\mu\nu} -{1\over 2} g_{\a\b} g^{\mu\nu}  \da^{(1)} R_{\mu\nu} =  \lp^2 \, \da^{(1)} T_{\a\b}, 
\label{230}
\eeq
where, by using Eq. (\ref{223}), 
\beq
g^{\mu\nu}  \da^{(1)} R_{\mu\nu}= \nabla_\mu \nabla_\nu h^{\mu\nu}- \nabla^2 h.
\label{231}
\eeq
It may be convenient, also, to rewrite the first two terms on the right-hand side of Eq. (\ref{223}) by applying the commutation rule of the covariant derivatives, which gives:
\beq
\nabla_\mu \nabla_\b h_\a\,^\mu= \nabla_\b \nabla_\mu h_\a\,^\mu +h_\a\,^\nu R_{\nu\b} - R_{\mu\a\b\nu}h^{\mu\nu},
\label{232}
\eeq
where $R_{\mu\a\b\nu}$ is the Riemann tensor. By combining all contributions, and multiplying by $-2$, we finally obtain the following dynamical evolution equation for the linear fluctuations $h_{\mu\nu}$ of a general background metric $g_{\mu\nu}$, with general matter sources $T_{\mu\nu}$, in the form:
\bea
&&~~~~
\nabla^2 h_{\a\b}+ 2 R_{\mu\a\b\nu} h^{\mu\nu}+ Rh_{\a\b} - g_{\a\b} h^{\mu\nu} R_{\mu\nu}- h_\a\,^\nu R_{\nu\b} - h_\b\,^\nu R_{\nu\a} - 
\nonumber \\ && \!\!\!\!\!\!
-\nabla_\b \nabla_\mu h_\a\,^\mu -\nabla_\a \nabla_\mu h_\b\,^\mu + \nabla_\a \nabla_\b h - g_{\a\b} (\nabla^2 h- \nabla_\mu\nabla_\nu h^{\mu\nu})= -2 \lp^2 \da^{(1)} T_{\a\b}. 
\label{233}
\eea
In a vacuum geometry ($T_{\mu\nu}=0= \da T_{\mu\nu}$, $R_{\mu\nu}=0=R$), and for metric fluctuations satisfying the so-called  TT gauge conditions, i.e.  $\nabla_\nu h_\mu\,^\nu=0= h_\mu\,^\mu$, we recover the well-known vacuum propagation equation  (see e.g. \cite{Mis}),
\beq
\nabla^2 h_{\a\b}+ 2 R_{\mu\a\b\nu} h^{\mu\nu}=0. 
\label{234}
\eeq

The general result (\ref{233}) -- for a non-vacuum geometry and with no gauge fixing -- was first derived (as far as we know) in a different but equivalent form in \cite{Sci}, and, in the same form as Eq. (\ref{233}), in an unpublished lecture note \cite{Det}. In both cases, however, it was derived without starting from a variational principle but directly perturbing, to first order, the background Einstein equations. It has been recently obtained with the same method  in \cite{Brev} (see also \cite{Fig}). In other papers (see e.g. \cite{Cap,Dol}) a similar equation is  presented by imposing however the TT conditions from the beginning, so that the last five terms on the left-hand side of Eq. (\ref{233}) are missing. It should be noted, in this regard, that in the presence of gravitational sources it is not possible, in general, to write the full metric perturbations in the TT gauge, as clearly stressed in \cite{Flan}. 

Let us conclude this Section by noting that the general propagation equation (\ref{233}) can also be rewritten in terms of another frequently used variable, the so called ``trace-reversed" perturbation $\psi_{\mu\nu}$, defined by:
\beq
\psi_{\mu\nu} \equiv h_{\mu\nu} -{1\over 2} g_{\mu\nu} h= h_{\mu\nu} +{1\over 2} g_{\mu\nu} \psi, ~~~~~~~~~~~~~
\psi \equiv g^{\mu\nu} \psi_{\mu\nu} = -h. 
\label{235}
\eeq
In terms of $\psi_{\mu\nu}$, our previous equation (\ref{233}) takes the form
\bea
&&
\nabla^2\psi_{\a\b}+ 2 R_{\mu\a\b\nu} \psi^{\mu\nu}+ R\,\psi_{\a\b} - g_{\a\b} \,\psi^{\mu\nu} R_{\mu\nu}- \psi_\a\,^\nu R_{\nu\b} - \psi_\b\,^\nu R_{\nu\a} - 
\nonumber \\ && 
-\nabla_\b \nabla_\mu\psi_\a\,^\mu -\nabla_\a \nabla_\mu \psi_\b\,^\mu +g_{\a\b}  \nabla_\mu\nabla_\nu \psi^{\mu\nu}= -2 \lp^2 \,\da^{(1)} T_{\a\b}
\label{236}
\eea
(useful, in vacuum, to impose the so-called Lorentz gauge condition $\nabla_\nu \psi_\mu\,^\nu=0$, see e.g. \cite{And}). 


\section{Evolution of transverse and traceless metric fluctuations}
\label{sec3}
\setcounter{equation}{0}

Let us now recall that, in general, not all the independent components of the metric perturbations $h_{\mu\nu}$ may satisfy the condition of vanishing trace and vanishing covariant divergence when the stress  tensor of the matter sources is nonzero.  In the rest of this paper, however, we will restrict our discussion to those metric perturbations (and to the associated background geometries) which satisfy such conditions, and which describe radiative degrees of freedom (GW) possibly propagating to infinity (as illustrated in details in \cite{Flan}). Why are we interested in such modes? Because we want to discuss the possible effects of the geometric  sources on the dynamics of  GW propagation, considering in particular  non-vacuum backgrounds of cosmological type, generated by conventional (or more ``exotic") classes of energy-momentum distributions. 

We shall thus consider components of the metric perturbations satisfying
 the conditions
\beq
\nabla _\nu h^{\mu\nu}=0, ~~~~~~~~~~~~  h= g^{\mu\nu}h_{\mu\nu}=0,
\label{31}
\eeq
which we shall call, for brevity, ``TT gauge" conditions. In that case  we have also $\psi=0$, $\psi_{\mu\nu}= h_{\mu\nu}$, and the corresponding propagation equation (\ref{233}) -- or, equivalently, our Eq. (\ref{236}) -- reduces to the simplified form
\beq
\nabla^2 h_{\a\b}+ 2 R_{\mu\a\b\nu} h^{\mu\nu}+ Rh_{\a\b} - g_{\a\b} \,h^{\mu\nu} R_{\mu\nu}- h_\a\,^\nu R_{\nu\b} - h_\b\,^\nu R_{\nu\a} - 
= -2 \lp^2 \,\da^{(1)} T_{\a\b}. 
\label{32}
\eeq
By computing the trace  (with respect to the unperturbed metric $g^{\a\b}$) of the above equation we then find, for consistency with the selected gauge,  that 
the background geometry and its sources must satisfy the condition
\beq
2 R_{\a\b} h^{\a\b} = \lp^2 \,g^{\a\b} \da^{(1)} T_{\a\b}.
\label{33}
\eeq
Note that for a vacuum, Ricci-flat metric ($T_{\mu\nu}=0=R_{\mu\nu}$), the above  equation is always automatically satisfied. 

It may be convenient, for the application of this paper, to rewrite Eq. (\ref{32}) in a form which better displays the possible modifications induced {by} the matter sources with respect to the standard GW propagation in vacuum (described by Eq. (\ref{234})). By using the background equations (\ref{219}) and the consistency condition (\ref{33}) 
 to eliminate everywhere the Ricci tensor, we can then rewrite Eq. (\ref{32}) as follows:
 \beq
\nabla^2 h_{\a\b}+ 2 R_{\mu\a\b\nu} h^{\mu\nu} = \lp^2 \left( h_\a\,^\nu T_{\nu\b}+ h_\b\,^\nu T_{\nu\a}+{1\over 2} g_{\a\b} \,g^{\mu\nu} \da^{(1)} T_{\mu\nu} -2 \da^{(1)} T_{\a\b}\right).
\label{34}
\eeq
Let us finally recall that, as previously stressed, the symbol $\da^{(1)} T_{\a\b}$ represents the second-order perturbation of the matter Lagrangian performed with respect to the metric only (see Eqs. (\ref{220}), (\ref{221})), and it does not imply any direct perturbation of the other specific and intrinsic  variables (like density distributions, velocity distributions) typical of the matter sources that we are considering.

In order to display the physical differences with vacuum GW propagation
it may be appropriate, at this point, to write down the modified perturbation equation (\ref{34}) explicitly,  for some specific case of non-vacuum background geometry satisfying the consistency condition (\ref{33}). To this purpose we shall provide  two examples (typical of cosmological applications),  where the sources of the unperturbed geometry are represented by a fluid or by a self-interacting scalar field. 

\subsection{Perfect fluid source}
\label{sec31}
 
 Let us consider a perfect fluid with barotropic equation of state, energy density $\r$, pressure $p$, four-velocity of the fluid element  $u^\mu = dx^\mu/ d\tau$, described by the Lagrangian density (see e.g. \cite{For}):
\beq
\sqrt{-g} \L =-{1\over 2} \sqrt{-g} \,\left[ \left(\r+p\right) g_{\r\sg} u^\r u^\sg -\left(\r+3p\right) \right].
\label{35}
\eeq
For the explicit application of Eqs. (\ref{33}) and (\ref{34}) we need both the 
energy-momentum tensor of the unperturbed fluid, given (according to Eq. (\ref{212})) by
\beq
T_{\a\b} = \left( {2\over \sqrt{-\og}} 
{\da \sqrt{-\og} \,\L \over \da \og^{\a\b}}\right)_0,
\label{36}
\eeq
and its first-order perturbation, given by
\beq
\da^{(1)} T_{\a\b} = \left( {\da \over \da \og_{\mu\nu}} {2\over \sqrt{-\og}} 
{\da \sqrt{-\og} \,\L \over \da \og^{\a\b}}\right)_0 \da g_{\mu\nu}.
\label{37}
\eeq

From the definition (\ref{36}), by using the metric property (\ref{213}) and imposing -- but only {\it after} the functional differentiation -- the standard  normalization  condition 
$ g_{\r\sg} u^\r u^\sg=1$, we are then led to the well known result
\beq
T_{\a\b}= - p\, g_{\a\b} + (\r+p) u_a u_\b.
\label{38}
\eeq
Let us now compute the second functional derivative, by applying the definition (\ref{37}) to the fluid Lagrangian (\ref{35}). We easily obtain:
\bea
\da^{(1)} T_{\a\b} &=&
\left( {\da \over \da \og_{\mu\nu}} \left[ {1\over 2} (\r+p) \og_{\a\b} \og_{\r\sg} u^\r u^\sg - {1\over 2} (\r+3p) \,\og_{\a\b} + (\r+p) \, u^\r u^\sg \,\og_{\r\a} \og_{\sg\b} \right] \right)_0 \da g_{\mu\nu}
\nonumber \\
&=&-p\, h_{\a\b} +{1\over 2} \left(\r+p\right) g_{\a\b}\, h_{\mu\nu} u^\mu u^\nu + (\r+p) \left(u_\a u^\nu h_{\nu\b} + u_\b u^\nu h_{\nu \a} \right). 
\label{39}
\eea
On the other hand, by using the background equations (\ref{219}), by inserting the stress tensor (\ref{38}) into the left left-hand side of the condition (\ref{33}), and taking into account that $h=0$, we obtain
\beq
2 R_{\a\b}h^{\a\b}= 2 \lp^2
\left(\r+p\right)u_\a u_\b h^{\a\b}.
\label{310}
\eeq
By inserting the perturbed stress-tensor (\ref{39}) into the right-hand side of Eq. (\ref{33}) we have:
\beq
 \lp^2 \,g^{\a\b} \da^{(1)} T_{\a\b} = 4 \lp^2
\left(\r+p\right) h_{\mu\nu} u^\mu u^\nu .
\label{311}
\eeq
It follows that the  condition of consistency with the TT gauge, for a perfect fluid source, is satisfied if and only if
\beq
\left(\r+p\right) h_{\mu\nu} u^\mu u^\nu =0
\label{312}
\eeq
(and in that case both sides of Eq. (\ref{33}) are identically vanishing). The above  constraint can be satisfied in two ways.

A first possibility is the case  $p=-\r$ (which describes, in a fluid dynamics language, a background geometry generated by a cosmological constant $\La= \r=$ cost). In that case we have $T_{\a\b}= - p g_{\a\b}$, $\da^{(1)} T_{\a\b}= -p h_{\a\b}$, and we can check that the right-hand side of Eq. (\ref{34}) is identically vanishing. It turns out that the linearized propagation equation of the metric fluctuations satisfying the TT gauge condition is exactly the same as that obtained in the context of a vacuum, Ricci-flat geometry, and is given by Eq. (\ref{234}). 

A second possibility to be consistent with the TT gauge is to assume that  $h_{\mu\nu} u^\mu u^\nu=0$. In that case the right-hand side of Eq. (\ref{34}) is in general non-vanishing,  and by using the explicit results (\ref{38}), (\ref{39}) for the fluid stress tensor we find:
\beq
\nabla^2 h_{\a\b}+ 2 R_{\mu\a\b\nu} h^{\mu\nu} +\lp^2 
(\r+p) \left(u_\a u^\nu h_{\nu\b} + u_\b u^\nu h_{\nu \a} \right)=0.
\label{313}
\eeq
Interestingly enough, we can then obtain a propagation equation different from the vacuum equation even for the components of the metric fluctuations which satisfy the TT gauge conditions, and which are typical of gravitational radiation.  

The above ``non standard" correction terms disappear, however,  from the previous equation for metric fluctuations satisfying the condition $h_{\mu\nu}u^\nu = 0$. For instance, they may disappear if the fluid has a ``comoving"  energy-momentum distribution, i.e. the fluid element has velocity $u^i=0$, $u^0=1$, and we assume
that the TT gauge is valid for the spatial components $h_{ij}$ of the metric fluctuations, with $h_{\mu0}=0$. For such  modes the consistency condition $h_{\mu\nu} u^\mu u^\nu=0$ is also automatically satisfied, and 
%
%
the propagation equation (\ref{313}) reduces to  
\beq
\nabla^2 h_{ij}+ 2 R_{kijl} h^{kl}=0.
\label{314}
\eeq
If we take a spatially flat, homogeneous and isotropic geometry,  described by a  FLRW metric with $g_{00}=1$, $g_{ij}=- a^2 \da_{ij}$, we have the following components of the connection and of the curvature tensor,
\beq
\Ga_{0i}\,^j= H \da_i^j, ~~~~~~~~~ \Ga_{ij}\,^0= - g_{ij}H, ~~~~~~~~~
R_{kijl} h^{kl} = H^2 h_{ij},
\label{315}
\eeq
where $H= \dot a/a$ and the dot denotes differentiation with respect to the cosmic time $t$. In such a background, if we explicitly write the propagation equation (\ref{314}) for the mixed components $h_i\,^j$ of the metric perturbations, $h_i\,^j\equiv g^{jk}h_{ik}$, where $\pa_j h_i\,^j=0=g^{jk} h_{jk}$,  
we then recover the standard, well-known result (see e.g. \cite{Muk})
\beq
\ddot h_i\,^j + 3 H \dot  h_i\,^j -{\pa^2\over a^2} \,h_i\,^j=0
\label{316}
\eeq
(where $\pa^2= \da^{kl} \pa_k\pa_l$).

\subsection{Minimally coupled scalar field}
\label{sec32}

Another typical source of the spacetime geometry at the cosmic level is possibly represented by a self-interacting scalar field $\phi$, described by the Lagrangian density:
\beq
\sqrt{-g} \,\L= \sqrt{-g} \left[{1\over 2} g^{\r\sg} \pa_\r \phi \pa_\sg \phi - V(\phi) \right]
\label{317}
\eeq
(which also includes the case of a cosmological constant $V=\La=$ cost). The corresponding energy-momentum tensor, defined by Eq. (\ref{36}), is given by
\beq
T_{\a\b}= \pa_\a \phi \pa_\b \phi - g_{\a\b} \left({1\over 2}  g^{\r\sg} \pa_\r \phi \pa_\sg \phi - V\right),
\label{318}
\eeq
and its first order perturbation, according to Eq. (\ref{37}), leads to
\beq
\da^{(1)} T_{\a\b}=h_{\a\b} \left(V- {1\over 2} \pa_\mu \phi \pa^\mu \phi \right) +{1\over 2} g_{\a\b}\, h^{\mu\nu} \pa_\mu \phi \pa_\nu \phi.
\label{319}
\eeq
By inserting the two above results into Eq. (\ref{33}), by using the background equations to eliminate $R_{\a\b}$ in terms of $T_{\a\b}$, and taking into account that $h=0$, we then find that the consistency condition with the TT gauge in this case is always automatically satisfied, namely:
\beq
2 h^{\a\b} R_{\a\b} \equiv 2 \lp^2 h^{\a\b} \pa_\a \phi \,\pa_\b \phi  \equiv
\lp^2 \, g^{\a\b} \da^{(1)} T_{\a\b},
\label{320}
\eeq
for any given type of scalar field dynamics. 

This does not imply, however, that there are no modifications to the vacuum propagation equation. By inserting into the right-hand side of Eq. (\ref{34}) the results (\ref{318}), (\ref{319}), we find 
\beq
\nabla^2 h_{\a\b}+ 2 R_{\mu\a\b\nu} h^{\mu\nu} -{\lp^2}  
 \left(h_\a\,^\nu \pa_\b \phi +  h_\b\,^\nu \pa_\a \phi \right) \pa_\nu \phi =0.
\label{321}
\eeq
If we have a homogeneous scalar source, with $\pa_i\phi=0$, it follows that  there are no corrections to the propagation equation for the spatial fluctuations  $h_{ij}$ (like in the case of the comoving fluid source). But non-trivial corrections are in general possible, even for the spatial fluctuations satisfying the TT gauge conditions, if  $h_\a\,^\nu \pa_\nu \phi \not=0$.

\subsection{Fluid source with viscosity}
\label{sec33}

Let us finally consider a possible source of the background geometry which can be described as a fluid with energy density $\r$, pressure $p$, velocity field $u^\mu$, bulk viscosity $\zeta$, and shear viscosity $\eta$, where $\zeta$ and $\eta$ are constant parameters. By introducing the effective pressure $\ti p$,
\beq
\ti p =p - \left(\zeta -{2\over 3} \eta \right) \nabla_\la u^\la,
\label{322}
\eeq
we can put the full energy-momentum tensor of the viscous fluid \cite{Mis,Wei,Gron}
in the following convenient form (see e.g. \cite{Ana,Mon} for recent discussions):
\beq
\ti T_{\a\b} = \left(\r+ \ti p \right) u_\a u_\b - \ti p g_{\a\b} - \eta \left[ u_\a u^\r \nabla_\r u_\b + u_\b u^\r \nabla_\r u_\a - \nabla_\a u_\b - \nabla_\b u_\a    \right].
\label{323}
\eeq
The first-order perturbation can then be written as
\bea 
\da^{(1)} \ti T_{\a\b} &=& \da^{(1)} T_{\a\b}(\ti p) + \left( u_\a u_\b - g_{\a\b} \right) 
 \left({2\over 3} \eta - \zeta \right) \da^{(1)} (\nabla_\r u^\r )
 \nonumber \\ &-&
\eta \, \da^{(1)} \left[ \left(g_{\a\r} g _{\b\sg} +  g_{\b\r} g _{\a\sg}\right)u^\r u^\la \nabla_\la u^\sg - g_{\b\r} \nabla_\a u^\r -  g_{\a\r} \nabla_\b u^\r
 \right],
\label{324}
\eea
where $ \da^{(1)} T_{\a\b}(\ti p)$ denotes the perfect fluid  result ({\ref{39}), written however in terms of the total effettive pressure $\ti p$.

By using the result (\ref{218}) for the first-order perturbation of the connection we find, in the TT gauge,\beq
 \da^{(1)} (\nabla_\r u^\r )= u^\a  \da^{(1)} \Gamma_{\nu \a}\,^\nu = {1\over 2} u^\a \nabla_\a (g^{\nu\b} h_{\nu\b}) \equiv {1\over 2} u^\a \nabla_\a h =0.
\label{325}
\eeq
so that the second contribution to Eq. (\ref{324}) is identically vanishing. Let us explicitly compute the remaining contributions, by using in particular the following results:
\bea
u_\a g_{\b\sg} u^\la  \da^{(1)} (\nabla_\la u^\sg ) &=& u_\a u^\la u^\r \left( \nabla_\la h_{\b\r} -{1\over 2} \nabla_\b h_{\la\r} \right),
\label{326}
\\ 
g_{\a\sg}  \da^{(1)} (\nabla_\b u^\sg )&=&{1\over 2} u^\r \left( \nabla_\b h_{\r\a} + \nabla_\r h_{\b\a}- \nabla_\a h_{\b\r} \right).
\label{327}
\eea
We finally obtain:
\bea 
\da^{(1)} \ti T_{\a\b} &=& \da^{(1)} T_{\a\b}(\ti p)
-\eta \left[ h_{\a\r} u^\r u^\la \nabla_\la u_\b + u_\a h_{\b\sg} u^\la \nabla_\la u^\sg -
 h_{\a\sg} \nabla_\b u^\sg 
\right.
 \nonumber\\ &+& 
 \left.
 u_\a u^\la u^\r \left( \nabla_\la h_{\b\r} -{1\over 2} \nabla_\b h_{\la\r} \right) 
+ ( \a \leftrightarrow \b) \right]  + \eta \, u^\la \nabla_\la h_{\a\b}, 
\label{328}
\eea
where the symbol $\{ \a \leftrightarrow \b\}$ denotes that all the preceding terms inside the square brackets are to be repeated with the index $\a$ replaced by $\b$ and vice-versa. 
We can now impose the consistency condition for the TT gauge. By computing the left-hand and right-hand side of Eq. (\ref{33}) for our viscous fluid source, and imposing the equality, we obtain the condition
\beq
2 \left(\r+ \ti p \right) h_{\mu\nu} u^\mu u^\nu = \eta \left[2 h^{\mu\nu} \nabla_\mu u_\nu + u^\b u^\la u^\r \nabla_\la h_{\b\r} \right].
\label{329}
\eeq
In the absence of shear viscosity ($\eta=0$), it simply reduces to the same consistency condition of the perfect fluid (\ref{312}) (with the possible contribution of bulk viscosity, contained inside the effective pressure $\ti p$). 

We are now in the position of discussing whether the presence of viscosity can modify or not the propagation of those metric fluctuations satisfying the TT gauge condition.  To this purpose, we have to explicitly evaluate the right-hand side of Eq. (\ref{34}) by using, for the viscous fluid, the results (\ref{323}) and (\ref{328}).
We obtain, in this way, a propagation equation which generalizes the perfect fluid equation (\ref{313}), and introduces many additional new terms depending on the viscosity parameters, on the covariant derivatives of the velocity $u^\mu$, and on the perturbation components  $h_{\a\b}$. For the purpose of this paper it will enough to illustrate the case, very simple (but physically relevant for cosmological applications), of a viscous fluid with comoving velocity distribution, $u^0=1$, $u^i=0$, which is source of a background geometry described by a  spatially flat FLRW metric (whose relevant connection and curvature components have been already reported in  Eq. (\ref{315})). 

In that case we can satisfy the consistency condition (\ref{329}) by assuming, as in the perfect fluid case, that $h_{0\mu}=0$, and that the TT gauge is valid for the spatial part $h_{ij}$ of the metric fluctuations. We then find that  the source contributions on the right-hand side of Eq. (\ref{34}) are all vanishing, with the only exception of the contribution arising from the last term of the perturbed stress tensor (\ref{328}), generated by the shear viscosity and proportional to 
\beq
u^\la \nabla_\la h_{ij} \equiv \dot h_{ij} - 2 H h_{ij}.
\label{330}
\eeq
By including this contribution, the propagation equation (\ref{34}) then takes the form
\beq
\nabla^2 h_{ij}+ 2 R_{kijl} h^{kl}+ 2 \eta\, \lp^2 \left(  \dot h_{ij} - 2 H h_{ij}  \right)=0.
\label{331}
\eeq
Written in terms of the mixed components, $h_i\,^k= g^{kj} h_{ij}$, we are finally lead to the result: 
\beq
\ddot h_i\,^j + \left(3 H+ 2 \eta\lp^2\right) \dot  h_i\,^j -{\pa^2\over a^2} \,h_i\,^j=0 
\label{332}
\eeq
(the same equation was previously obtained, but with a different procedure, also in \cite{Pra}). 
For vanishing (or negligible) viscosity, $\eta \ra 0$,  we recover the  standard equation  (\ref{316}). It is important to note, however, that for $\eta \lp^2 \sim H$ the presence of shear viscosity can  affect in a significant way the evolution and the amplification of  tensor metric fluctuations, and  the consequent production of a relic cosmic GW background. 

It may be useful to note, finally,  that the above equation can be rewritten in terms of the conformal time coordinate $\tau$ (related to the cosmic time $t$ by $dt= a d\tau$) as follows:
\beq
 h''_i\,^j +2\left( {\cal H}+ \eta \lp^2 a\right)  h'_i\,^j -\pa^2  h_i\,^j=0,
\label{333}
\eeq
where ${\cal H}=a'/a$, and the prime denotes differentiation with respect to $\tau$. 
In other words, we can say that the GW dynamics is affected by a friction coefficient 
 $\da$ such that:
\beq
 h''_i\,^j +2{\cal H}h'_i\,^j \left[1-\da(\tau)\right] -\pa^2 h_i\,^j=0,
\label{334}
\eeq
where $\da(\tau) =- \eta \lp^2 a/{\cal H}=-\eta \lp^2 a^2/a'$. It follows that the presence of shear viscosity, correctly included into the propagation dynamics of tensor metric perturbations according to Eq. (\ref{34}), 
may produce effects very similar to those obtained in the context of modified theories of gravity (see e.g. the models and the examples discussed in 
\cite{Mag,Fan}).

\section{Conclusion}
\label{sec4}
\setcounter{equation}{0}

In this paper we have presented a variational method to obtain the equations governing the evolution of linear metric perturbations in a general background geometry, with general sources and general choice of the coordinates. We have considered in particular the Einstein model of gravity, but the method can be applied to any given action assumed to describe the gravitational dynamics. Also, we have applied the action variational formalism to the metric variable only, without perturbing other field variables typical of the matter sources. 

We have concentrated our discussion on the metric fluctuations satisfying the transversality and traceless conditions, and on the associated background geometries compatible with such conditions. In that case, we have derived a general equation describing the propagation of gravitational radiation, and the important result is that the effective form of such equation depends (as expected) not only on the background geometry, but also -- for the same given geometry -- on {\it how} the metric couples to the gravitational sources inside the matter part of the action. 

We have given explicit examples for standard cosmological geometries generated by fluid sources, with and without bulk and shear viscosity, and we have found that the metric coupling to the shear viscosity is able in principle to leave an imprint on the primordial GW spectrum (an effect that we plan to discuss in detail in a future  paper). The case of electromagnetic field sources, presented in the Appendix B, is also worth of further study. Finally, we are planning to apply the method of this paper to generalize the results obtained for metric perturbations in less trivial cosmological scenarios, 
like those described by background geometries of Bianchi-type,  Lema\^itre-Tolman-Bondy-type, or Light-Cone-type 
(see e.g. \cite{Pereira,Fanizza2,Fanizza1,Mitsou}).

\section*{Acknowledgements}

Maurizio Gasperini and Luigi Tedesco are supported in part by INFN under the program TAsP ({\it ``Theoretical Astroparticle Physics"}), and by the research grant number 2017W4HA7S {\it ``NAT-NET: Neutrino and Astroparticle Theory Network"}, under the program PRIN 2017 funded by the Italian Ministero dell'Universit\`a e della Ricerca (MUR). Giuseppe Fanizza acknowledges support by FCT under the program {\it ``Stimulus"} with the grant CEECIND/04399/\\2017/CP1387/CT0026. It is a pleasure to thank Gabriele Veneziano for useful comments and discussions. 


\appendix
\renewcommand{\theequation}{A.\arabic{equation}}
\setcounter{equation}{0}
\section{Appendix A. Second-order expansion of the Christoffel connection and of the Ricci tensor}

Let us start by deriving the result of Eq. (\ref{227}), namely the second-order term in the expansion of the Christoffel connection in series of functional derivatives with respect to the metric, under the action of the infinitesimal transformation (\ref{21}). 

Using the definition
\beq
\Ga_{\nu\a}\,^\mu= {1\over 2} g^{\mu\b} \left(\pa_\nu g_{\a\b}+\pa_\a g_{\nu\b} - \pa_\b g_{\nu\a} \right),
\label{a1}
\eeq
we find that the first functional derivative, by applying  Eqs. (\ref{23}) and (\ref{25}), is given by:
\bea
\da^{(1)} \Ga_{\nu\a}\,^\mu = &-&{1\over 2}g^{\mu\r} g^{\b\sg} \left(\pa_\nu g_{\a\b}+\pa_\a g_{\nu\b} - \pa_\b g_{\nu\a} \right)\da g_{\r\sg}
\nonumber \\
&+& {1\over 2}g^{\mu\b}\left( \da^{\la\r\sg}_{\nu\a\b}+  \da^{\la\r\sg}_{\a\nu\b}
- \da^{\la\r\sg}_{\b\nu\a}\right) \pa_\la \da g_{\r\sg},
\label{a2}
\eea
where $ \da^{\la\r\sg}_{\nu\a\b}\equiv \da^\la_\nu \da^\r_\a \da^\sg_\b$. A second application of  the functional derivative operator then gives
\bea
\da^{(2)} \Ga_{\nu\a}\,^\mu &=&{1\over 2} {\da\over \da g_{\ga\da}} \left(\da^{(1)} \Ga_{\nu\a}\,^\mu \right) \da g_{\ga\da}=
\nonumber \\
&=&{1\over 4}\left(g^{\mu\ga}g^{\r\da}g^{\b\sg} +g^{\mu\r}g^{\b\ga}g^{\sg\da}\right)
\left(\pa_\nu g_{\a\b}+\pa_\a g_{\nu\b} - \pa_\b g_{\nu\a} \right)  \da g_{\ga\da}\,
 \da g_{\r\sg}
 \nonumber \\
&-&{1\over 4}g^{\mu\r} g^{\b\sg} \left( \da^{\la\ga\da}_{\nu\a\b}+  \da^{\la\ga\da}_{\a\nu\b} - \da^{\la\ga\da}_{\b\nu\a}\right) (\pa_\la \da g_{\ga\da})\, \da g_{\r\sg}
 \nonumber \\
&-&{1\over 4}\left( g^{\mu\ga} g^{\sg\da} \da^{\la\r}_{\nu\a}+ g^{\mu\ga} g^{\sg\da} \da^{\la\r}_{\a\nu}- g^{\mu\ga} g^{\la\da} \da^{\r\sg}_{\nu\a} 
\right)\da g_{\ga\da}\,\pa_\la \da g_{\r\sg},
\label{a3}
\eea
which implies, after a few simple steps of tensor algebra, 
\bea
\da^{(2)} \Ga_{\nu\a}\,^\mu &=&{1\over 2} h^{\mu\la}h_\la\,^\b \left(\pa_\nu g_{\a\b}+\pa_\a g_{\nu\b} - \pa_\b g_{\nu\a} \right) 
 \nonumber \\
&-&{1\over 2}h^{\mu\b}\left(\pa_\nu h_{\a\b}+\pa_\a h_{\nu\b} - \pa_\b h_{\nu\a} \right) .
\label{a4}
\eea
Let us finally use the identity
\beq
\pa_\nu h_{\a\b}+\pa_\a h_{\nu\b} - \pa_\b h_{\nu\a}  =
\nabla_\nu h_{\a\b}+\nabla_\a h_{\nu\b} - \nabla_\b h_{\nu\a} +2 \Ga_{\nu\a}\,^\r h_{\r\b}, 
\label{a5}
\eeq
which, inserted into Eq. (\ref{a4}), leads to the final result of Eq. (\ref{227}):
\beq
\da^{(2)} \Ga_{\nu\a}\,^\mu =  -{1\over 2}  h^{\mu\b} \left( \nabla_\nu h_{\a\b}
+\nabla_\a h_{\nu\b}-\nabla_\b h_{\nu\a} \right).
\label{a6}
\eeq

Let us now compute the corresponding expansion of the Ricci tensor up to terms quadratic in $\da g_{\mu\nu}$. Starting from the definition
\beq
R_{\nu\a} = \pa _\mu \Ga _{\nu\a}\,^\mu -  \pa _\nu \Ga _{\mu\a}\,^\mu + \Ga_{\mu\r}\,^\mu \Ga_{\nu\a}\,^\r- \Ga_{\nu\r}\,^\mu \Ga_{\mu\a}\,^\r,
\label{a7}
\eeq
we have
\beq
\da^{(2)} R_{\nu\a} = \pa _\mu \da^{(2)} \Ga _{\nu\a}\,^\mu + 
\Ga_{\mu\r}\,^\mu \da^{(2)}\Ga_{\nu\a}\,^\r + \da^{(1)}\Ga_{\mu\r}\,^\mu \da^{(1)}\Ga_{\nu\a}\,^\r +\da^{(2)}\Ga_{\mu\r}\,^\mu \Ga_{\nu\a}\,^\r
- \{ \mu \leftrightarrow \nu\},
\label{a8}
\eeq
where the symbol $\{ \mu \leftrightarrow \nu\}$ denotes that all the preceding terms are to be repeated with the index $\mu$ replaced by $\nu$ and vice-versa. By using the definition of covariant derivative, 
\beq
\nabla_\mu \left(\da^{(2)} \Ga _{\nu\a}\,^\mu\right) =  \pa _\mu \left(\da^{(2)} \Ga _{\nu\a}\,^\mu\right) -
\Ga_{\mu\nu}\,^\r \da^{(2)} \Ga _{\r\a}\,^\mu - \Ga_{\mu\a}\,^\r \da^{(2)} \Ga _{\nu\r}\,^\mu +\Ga_{\mu\r}\,^\mu \da^{(2)} \Ga _{\nu\a}\,^\r , 
\label{a9}
\eeq
we can  eliminate in Eq. (\ref{a8}) the partial derivatives of the perturbed connection in terms of the corresponding covariant ones. We then find that all contributions of the type $\Ga  \da^{(2)} \Ga$ present in Eq. (\ref{a8}) either cancel or combine to contribute to the covariant derivatives, while the quadratic  terms of the type  $ \da^{(1)} \Ga  \da^{(1)} \Ga$  keep unchanged. We then obtain
\beq
\da^{(2)} R_{\nu\a} = \nabla _\mu \left(\da^{(2)} \Ga _{\nu\a}\,^\mu \right) + \da^{(1)}\Ga_{\mu\r}\,^\mu \da^{(1)}\Ga_{\nu\a}\,^\r - \{ \mu \leftrightarrow \nu\},
\label{a10}
\eeq
which exactly reproduces the result (\ref{225}) reported in Sect. \ref{sec22}. 


\renewcommand{\theequation}{B.\arabic{equation}}
\setcounter{equation}{0}
\section{Appendix B.  Electromagnetic fields}

Results similar to the ones of the perfect fluid case (see Sec. \ref{sec31}) can be obtained by considering an electromagnetic field $F_{\mu\nu}$ as the source of the background geometry, and starting with the Lagrangian density
\beq
\sqrt{-g} \L =-{\sqrt{-g}\over 16 \pi} g^{\r\la} g^{\sg \da} F_{\r\sg} F_{\la\da}.
\label{b1}
\eeq
The definition (\ref{36}) leads then the energy-momentum tensor
\beq
T_{\a\b}= -{1\over 4 \pi} \left(F_{\a\la} F_\b\,^\la - {1\over 4} g_{\a\b} F^{\mu\nu} F_{\mu\nu}\right),
\label{b2}
\eeq
and its perturbation gives:
\beq
 \da^{(1)} T_{\a\b} = {1\over 4 \pi} \left(F_{\a\mu} F_{\b\nu} h^{\mu\nu} +{1\over 4} h_{\a\b}  F^{\mu\nu} F_{\mu\nu} -{1\over 2} g_{\a\b}F_{\mu\la} F_\nu\,^\la h^{\mu\nu} \right).
 \label{b3}
 \eeq
The stress tensor (\ref{b2}) is traceless, the corresponding Ricci tensor is simply given by $R_{\a\b}= \lp^2 T_{\a\b}$, and the left-hand side of eq. (\ref{33}), with the condition $h=0$, can be written as
\beq
2 R_{\a\b} h^{\a\b} = - {\lp^2 \over 2 \pi} F_{\a\la} F_\b\,^\la h^{\a\b}.
\label{b4}
\eeq
By inserting the perturbed stress tensor (\ref{b3}) on the right-hand side of Eq. (\ref{33}) (again, for a traceless perturbation) we obtain:
\beq
\lp^2 g^{\a\b}  \da^{(1)} T_{\a\b} = - {\lp^2 \over 4 \pi} F_{\mu\la} F_\nu\,^\la h^{\mu\nu}.
\label{b5}
\eeq
The consistency condition for the TT gauge is thus satisfied, for the electromagnetic field sources, if and only if
\beq
F_{\mu\la}F_\nu\,^\la h^{\mu\nu}=0
\label{b6}
\eeq
(and in that case both sides of Eq. (\ref{33}) are identically vanishing). Again, we find that the possible existence of metric fluctuations satisfying the TT gauge may depend on the specific properties of the sources generating the background geometry.

Assuming that the TT gauge conditions are satisfied, let us finally discuss if (and how) the propagation dynamics can be affected by the electromagnetic field. By computing the right-hand side of Eq. (\ref{34}), and using the condition (\ref{b6}), we obtain the equation
\beq
\nabla^2 h_{\a\b}+ 2 R_{\mu\a\b\nu} h^{\mu\nu} +{\lp^2\over 4 \pi}  
 \left(h_\a\,^\nu F_{\nu\la} F_\b \,^\la + h_\b\,^\nu F_{\nu\la} F_\a\,^\la
 +2 F_{\a\mu} F_{\b\nu} h^{\mu\nu} \right)=0.
\label{b7}
\eeq
Hence, it is possible in principle to have a modified propagation dynamics, even for modes $h_\a\,^\nu$ satisfying the TT gauge, unless the background electromagnetic field $F_{\mu\nu}$ satisfies the condition $h_\a\,^\nu F_{\nu\la} =0$.



\end{document}